\documentclass[a4paper,11pt]{article}
\usepackage{pos}
\usepackage{orcidlink}

\title{Thermodynamic Consistency as a Reliability Test for Complex Langevin Simulations }
\ShortTitle{Thermodynamic Consistency as a Reliability Test for Complex Langevin Simulations }

\author*[a]{Anosh Joseph~\orcidlink{0000-0003-4288-8207}}
\affiliation[a]{National Institute for Theoretical and Computational Sciences, \\ School of Physics, and Mandelstam Institute for Theoretical Physics,\\ University of the Witwatersrand, Johannesburg, Wits 2050, South Africa}
\emailAdd{anosh.joseph@wits.ac.za}

\author[b]{Arpith Kumar~\orcidlink{0000-0002-5887-3803}}
\affiliation[b]{Key Laboratory of Quark and Lepton Physics (MOE) and Institute of Particle Physics, \\
Central China Normal University, Wuhan 430079, China}
\emailAdd{arpithk@ccnu.edu.cn}

\FullConference{The 42nd International Symposium on Lattice Field Theory (LATTICE2025)\\
2-8 November 2025\\
Tata Institute of Fundamental Research, Mumbai, India\\}

\abstract{
The complex Langevin method (CLM) is a promising tool to address the sign problem in quantum field theories with complex actions. However, it can converge to incorrect results even when simulations appear stable, highlighting the need for robust diagnostics. Existing checks, such as monitoring drift distributions, are useful but indirect. We propose a complementary test based on the configurational temperature, constructed from the gradient and Hessian of the complex action. Unlike drift-based criteria, this estimator directly probes thermodynamic consistency and provides a physically interpretable cross-check of CLM dynamics. Using one-dimensional PT-symmetric models, we show that it reproduces the input temperature with high precision and sensitively detects algorithmic errors, step-size artifacts, and incomplete thermalization. While demonstrated in simple systems, the method extends naturally to higher-dimensional scalar and gauge theories. Since temperature is tied to the bare coupling in many lattice theories, configurational monitoring can also provide an independent check on coupling-dependent observables. Our results indicate that configurational temperature can enhance CLM reliability across a broad range of applications, including lattice QCD at finite density.
}

\begin{document}
\maketitle

\section{Introduction}
\label{sec:intro}

Lattice regularizations of the path integral provide a systematic framework for studying the nonperturbative dynamics of quantum field theories (QFTs). 
In this approach, Monte Carlo methods are used to generate field configurations with probabilities proportional to the Euclidean weight $e^{-{\cal S}}$, allowing physical observables to be computed by importance sampling.
This strategy breaks down when the action becomes complex, since the weight can no longer be interpreted as a probability measure, leading to the well-known {\it sign problem}.
Such complex actions arise in many physically important settings, including QCD at finite baryon density, Chern--Simons theories with complex couplings, and certain chiral gauge theories, severely limiting the applicability of standard Monte Carlo techniques.

The complex Langevin method (CLM)~\cite{Klauder:1983nn, Parisi:1984cs} was introduced as a possible way to circumvent the sign problem by extending stochastic quantization to complex actions.
The method complexifies the original dynamical variables and evolves them according to a Langevin equation with complex drift.
Expectation values in the original path integral are then obtained as equilibrium averages over the resulting stochastic process.
For a recent review, see~\cite{Joseph:2025tfw}.

CLM has been successfully applied to a wide range of systems, including relativistic Bose gases at finite chemical potential, QCD-like theories in low dimensions, spin models at nonzero density, supersymmetric matrix models~\cite{Anagnostopoulos:2017gos, Joseph:2019sof, Joseph:2020gdh, Kumar:2022fas, Kumar:2022giw, Kumar:2023nya}, and large-$N$ unitary matrix models exhibiting Gross--Witten--Wadia transitions~\cite{Basu:2018dtm}.
Extending these successes to lattice QCD at finite density remains a central long-term goal, but doing so requires robust diagnostics to ensure that CLM converges to the correct results.

It is well known that CLM can converge to incorrect limits even when simulations appear stable.
This has motivated the development of reliability criteria based on, for example, the distribution of the drift term or properties of the Langevin-time evolution operator.
While valuable, such diagnostics can be insufficient in higher-dimensional systems or in the presence of subtle algorithmic pathologies.
In this work, we propose a complementary diagnostic based on configurational temperature that provides a direct test of thermodynamic consistency.

\section{Complex Langevin Method and Its Limitations}
\label{sec:Complex_Langevin_Method_and_Its_Limitations}

For a generic field $\phi$, we can write the complex Langevin equation in Euler-discretized form as
\begin{equation}
\phi(\theta + \Delta \theta) = \phi(\theta) - \Delta \theta \, \frac{\delta {\cal S}[\phi]}{\delta \phi(\theta)} + \sqrt{\Delta\theta} \, \eta(\theta).
\end{equation}
Here, $\theta$ denotes the Langevin time and $\eta(\theta)$ is Gaussian noise satisfying $\langle \eta(\theta) \rangle = 0$, $\langle \eta(\theta) \eta(\theta') \rangle = 2 \, \delta({\theta- \theta'})$. 
In practice, we use a real noise to suppress large excursions into the imaginary directions of the complexified fields. For an observable $\mathcal{O}$, we can write the noise-averaged expectation value as
\begin{equation}
\big\langle \mathcal{O}[\phi(\theta)] \big\rangle_\eta = \int d\phi \, P[\phi(\theta)] \, \mathcal{O}[\phi].
\end{equation}
The probability distribution $P[\phi(\theta)]$ evolves according to the associated Fokker--Planck equation.
For real actions, the stationary solution is $P \propto e^{-{\cal S}}$, ensuring convergence to the correct equilibrium distribution.  
For complex actions, however, the drift term is complex and the fields evolve on a complexified manifold, and a general proof of convergence to the desired complex measure is lacking.

It is well established that CLM can converge to incorrect limits despite apparently stable dynamics.
Two primary mechanisms have been identified.
The {\it excursion problem} arises when fields drift far into imaginary directions, invalidating the integration-by-parts arguments underlying the method.
The {\it singular drift problem} occurs when poles in the drift term are frequently sampled, again spoiling the justification of correctness.
These issues explain why CLM may yield systematically wrong results even in well-behaved simulations.

This has motivated the development of practical correctness criteria.
One proposal~\cite{Aarts:2009uq} requires the Langevin-time evolution operator $L$ to satisfy $\langle L\mathcal{O} \rangle = 0$ in the long-time limit.
This is conceptually appealing, however, this condition is difficult to verify numerically in large systems due to substantial statistical noise.
Nagata {\it et al.}~\cite{Nagata:2016vkn} formulated a more stringent criterion; they showed that correct convergence requires the probability distribution of the drift term to decay exponentially (or faster) at large magnitude.
This condition, under additional assumptions such as ergodicity, is both necessary and sufficient, and is straightforward to monitor during simulations. 
See Ref. \cite{Mandl:2025ins} for an update on recent developments regarding the use of kernels in complex Langevin simulations, to solve the problem of wrong convergence.

Nevertheless, existing diagnostics may be difficult to apply or interpret in complex, higher-dimensional theories.
This motivates the search for a complementary criterion tied more directly to physical consistency.
In this work, we propose such a diagnostic based on a configurational temperature estimator constructed from the gradient and Hessian of the complex Euclidean action.

\section{Configuration-Based Thermometer}
\label{sec:Configuration-Based_Thermometer}

A geometric definition of temperature was first introduced by Rugh in the microcanonical ensemble, where the inverse temperature is related to the curvature of constant-energy hypersurfaces in phase space~\cite{PhysRevLett.78.772}. 
This idea was later extended to canonical ensembles by Butler {\it et al.}, who formulated a \emph{configurational temperature} depending only on gradients and curvatures of the potential~\cite{10.1063/1.477301}. 
Because it requires no momenta, this estimator is particularly well suited for Monte Carlo and Langevin simulations.

\subsection{Configuration-based temperature estimator}
\label{sec:A_brief_derivation_of_the_configuration-based_thermometer}

The inverse temperature is defined thermodynamically as $1/T = (\partial S/\partial E) |_V$. In the microcanonical ensemble, the entropy is given by
\begin{equation}
S(E) = k_B \ln \Omega_\Gamma(E), \qquad \Omega_\Gamma(E) = \int_{\mu C(E)} d\vec{\Gamma},
\end{equation}
where $\vec{\Gamma} = (\vec{q}, \vec{p})$ and $\mu C(E) = \{\vec{\Gamma}\mid H(\vec{\Gamma}) \le E\}$.

Following Rugh, one considers an infinitesimal displacement along a phase-space vector field $\vec{n}(\vec{\Gamma})$ that increases the energy by $\Delta E$.  
Choosing $\vec{n} = \vec{\nabla}_{\vec{q}} H / (\vec{\nabla}_{\vec{q}} H \cdot \vec{\nabla}_{\vec{q}} H)$ ensures $H(\vec{q} + \Delta E \, \vec{n}) = H(\vec{q}) + \Delta E$ to leading order.  
The resulting Jacobian yields $1/ T = k_B \big\langle \vec{\nabla}_{\vec{q}} \cdot \vec{n} \big\rangle$.

For systems without explicit momenta, one sets $H(\vec{q}) = \Phi(\vec{q})$, leading to the configurational estimator
\begin{equation}
\frac{1}{k_B T} = \left\langle \vec{\nabla}_{\vec{q}} \cdot \frac{\vec{\nabla}_{\vec{q}} \Phi}{|\vec{\nabla}_{\vec{q}} \Phi|^2} \right\rangle + \mathcal{O}(1/N_{\rm dof}),
\label{eq:temp_eq}
\end{equation}
which is valid in the canonical ensemble by ensemble equivalence in the thermodynamic limit. In the above, $\Phi(\vec{q})$ denotes the potential and $N_{\rm dof}$ the number of degrees of freedom.

Writing $\vec{g} = \vec{\nabla}_{\vec{q}}\Phi$ and $\mathbb{H} = \vec{\nabla}_{\vec{q}} \vec{\nabla}_{\vec{q}}^T \Phi$, this can be expressed as
\begin{equation}
\frac{1}{k_B T} = \frac{\mathrm{Tr}(\mathbb{H})}{|\vec{g}|^2} - 2 \, \frac{\vec{g}^T\mathbb{H}\vec{g}}{|\vec{g}|^4}.
\label{eq:hessian-form}
\end{equation}

\subsection{Application to Euclidean lattice field theory}
\label{sec:Application_to_Euclidean_lattice_field_theory}

The Euclidean path integral has the same mathematical structure as a canonical ensemble,
\begin{equation}
\langle O \rangle = \frac{\int \mathcal{D} \phi \, O[\phi] \, e^{-{\cal S}[\phi]}} {\int \mathcal{D}\phi \, e^{- {\cal S}[\phi]}},
\end{equation}
with the action ${\cal S}[\phi]$ playing the role of a potential and inverse temperature fixed to unity.
We can use this formal analogy to motivate the use of the configurational temperature estimator in lattice field theory.

In this context, the configurational temperature does \emph{not} represent a physical thermodynamic temperature.
Instead, it serves as a diagnostic of sampling consistency.
Suppose an algorithm samples configurations with the weight $e^{- \alpha {\cal S}[\phi]}$ instead of the target $e^{- {\cal S}[\phi]}$.
Then the estimator will return $\beta_{\mathrm{conf}} = \alpha^{-1} \beta_{\mathrm{input}}$, directly revealing the deviation.

For lattice field theories, the estimator takes the form
\begin{equation}
\beta_{\mathrm{conf}} = \Bigg\langle \vec{\nabla}_\phi \cdot \frac{\vec{\nabla}_\phi {\cal S}[\phi]}{|\vec{\nabla}_\phi {\cal S}[\phi]|^2} \Bigg\rangle,
\label{eq:lattice_estimator}
\end{equation}
where derivatives are taken with respect to the lattice fields at each site. 
Comparing $\beta_{\mathrm{conf}}$ with the input inverse temperature provides a stringent test of algorithmic correctness, numerical stability, and thermalization.

\section{One-dimensional PT-symmetric theories}
\label{sec:One-dimensional_PT-symmetric_theories}

Quantum-mechanical theories with PT-symmetric, non-Hermitian Hamiltonians are known to possess real, positive spectra.
A simple example is the $(0+1)$D scalar theory with potential
\begin{equation}
\label{eqn:lat-scalar-pt-symm-pot}
V(\phi) = - \frac{g}{N} (i \phi)^N,
\end{equation}
where $g > 0$ and $N = 2 + \delta$ with $\delta > 0$.
The Euclidean action is defined on a thermal circle of circumference $\beta = 1 / T$,
\begin{equation}
{\cal S} = \int_0^\beta d\tau \left[ \tfrac{1}{2} \left( \tfrac{\partial\phi}{\partial\tau} \right)^2 + V(\phi) \right].
\end{equation}

Discretizing the theory on a lattice with $N_\tau$ sites we get the lattice action
\begin{equation}
{\cal S} = \sum_{n = 0}^{N_\tau - 1} \left[ \frac{(\phi_{n+1} - \phi_n)^2}{2} - \frac{g}{2 + \delta} (i \phi_n)^{2 + \delta} \right],
\end{equation}
with periodic boundary conditions for the scalar field.
The dimensionless field is related to the physical one by $\phi = \phi_{\rm phys} / \sqrt{a}$, the coupling scales as $g = a^{2 + \delta/2} g_{\rm phys}$, and $\beta = N_\tau a$.

We compute equal-time correlators $G_k \equiv \langle \phi^k \rangle$, $k = 1, 2$, for $\delta = 1, 2$.
Simulations were performed with $N_\tau = 128$, $a = 1$, adaptive step size $\epsilon \le 10^{-3}$, $N_{\rm therm} = 10^4$, and $N_{\rm gen} = 10^6$, $g = 1$, starting from $\phi_{\rm initial} = - i$.
Measurements were taken every 10 steps. 
For $\delta = 1$ we find $G_1 = \langle \phi \rangle = - i \, 0.5994(07)$, $G_2 = \langle \phi^2 \rangle = i \, 0.0033(43)$, while for $\delta = 2$, $G_1 = - i \, 0.8997(06)$ and $G_2 = - 0.5545(24)$, in agreement with previous studies, confirming correct sampling of the PT-symmetric stationary distribution.

\subsection{Temperature estimator}
\label{sec:Temperature_estimator}

We now evaluate the configurational temperature estimator in this model.
For each Langevin configuration $i$, we define
\begin{equation}
\hat{\beta}_i = \frac{\sum_n h_{nn}}{\sum_n g_n^2} - \frac{2 \sum_{nm} g_n g_m h_{nm}} {\left( \sum_n g_n^2 \right)^2}, \qquad g_n \equiv \frac{1}{\beta} \frac{\partial{\cal S}}{\partial\phi_n},
\qquad
h_{nm} \equiv \frac{1}{\beta} \frac{\partial^2{\cal S}}{\partial\phi_n\partial\phi_m}.
\end{equation}

The measured configurational temperature is defined as
\begin{equation}
\beta_M \equiv \mathrm{Re}\!\left[ \frac{1}{N_{\rm config}} \sum_{i = 1}^{N_{\rm config}} \hat{\beta}_i \right],
\end{equation}
where the imaginary part is consistent with zero.

\begin{figure*}[htbp]
\centering
\includegraphics[width=2.5in]{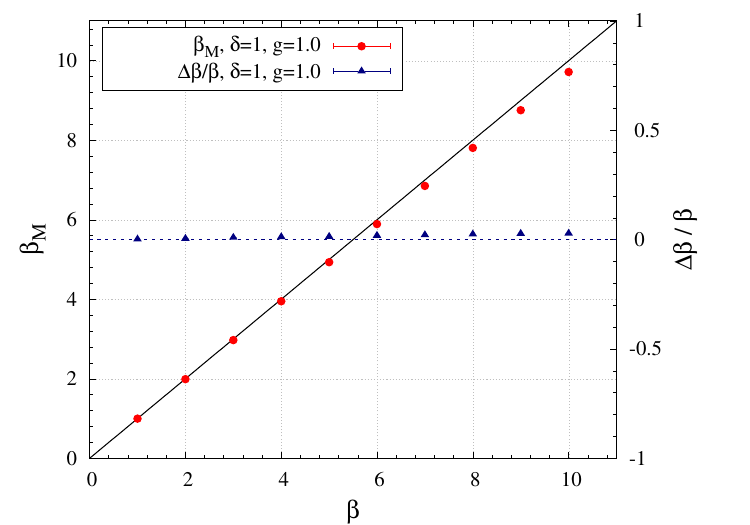}	
\includegraphics[width=2.5in]{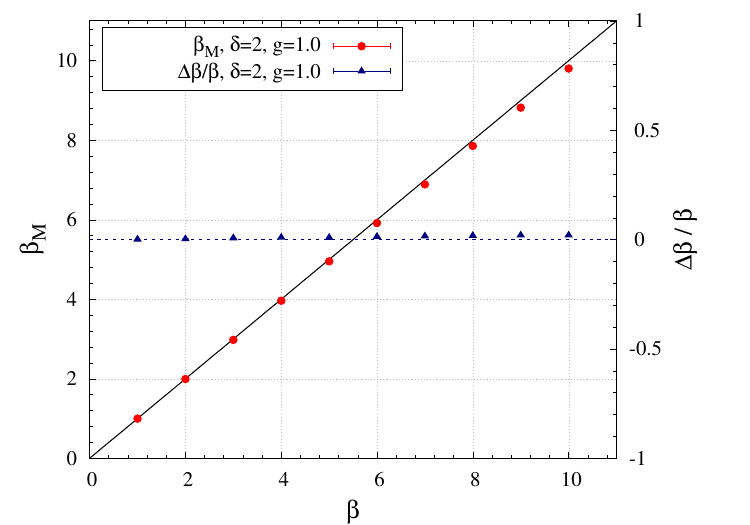}
\caption {The estimated configurational temperature $\beta_M$ and the error in estimated temperature, $(\beta - \beta_M)/\beta$ as functions of input $\beta$.}
\label{fig:beta_beta_m_delta}
\end{figure*}

Table~\ref{tab:lat-ptsymm-beta-n3-n4} compares $\beta_M$ with the input $\beta$ for $\delta = 1, 2$, while Fig.~\ref{fig:beta_beta_m_delta} shows $\beta_M$ and the relative deviation $(\beta - \beta_M) / \beta$ as functions of $\beta$.
The estimator reproduces the input temperature to within $0.2$ -- $3\%$ over the full range studied.
The deviation increases with $\beta$, reaching $\sim 2.8\%$ at $\beta = 10$, consistent with the expected $\mathcal{O}(1/N_\tau)$ finite-size corrections in Eq.~\eqref{eq:temp_eq}.
For $N_\tau = 128$, such corrections are naturally at the percent level, highlighting the sensitivity of the estimator to lattice artifacts.

\begin{table}[htbp]
\centering
\setlength{\tabcolsep}{4pt}
\renewcommand{\arraystretch}{1.1}
\begin{tabular}{c c c c c}
\hline
$\beta$ &
$\beta_M (\delta = 1)$ &
$\Delta\beta/\beta \, (\%)$ &
$\beta_M (\delta = 2)$ &
$\Delta\beta/\beta \, (\%)$ \\
\hline \hline
$1.00$  & $0.9983(14)$  & $0.17$ & $0.9994(22)$  & $0.06$ \\
$2.00$  & $1.9907(29)$  & $0.47$ & $1.9947(44)$  & $0.27$ \\
$5.00$  & $4.9321(72)$  & $1.36$ & $4.9549(109)$ & $0.90$ \\
$10.00$ & $9.7156(143)$ & $2.84$ & $9.8038(215)$ & $1.96$ \\
\hline
\end{tabular}
\caption{The input $\beta$ and estimated configurational temperature $\beta_M$ for the PT-symmetric model at $g = 1.0$.}
\label{tab:lat-ptsymm-beta-n3-n4}
\end{table}

\section{Numerical Tests of the Estimator}
\label{sec:Numerical_Tests_of_the_Estimator}

We can use the configurational temperature estimator to detect algorithmic errors, discretization artifacts, and equilibration effects.

\subsection{Detecting algorithmic errors}
\label{sec:Detecting_algorithmic_errors}

We introduce a controlled error by modifying the normalization of the noise in the Langevin equation, $\langle \eta(\theta) \eta(\theta') \rangle = \sigma \, \delta({\theta-\theta'})$, with $\sigma\neq 2$, instead of the correct normalization.  
This breaks the fluctuation--dissipation relation and changes the stationary distribution from $e^{- {\cal S}}$ to $e^{- (2 / \sigma){\cal S}}$.

As shown in Table~\ref{tab:lat_ptsymm_beta_algo_err}, the configurational estimator detects this immediately: $\beta_M$ shifts away from the input value $\beta$ in direct proportion to $\sigma$.  
We see that even mild mis-scalings (e.g., $\sigma = 1$) result in clear deviations, demonstrating that the estimator provides a sensitive and robust diagnostic of algorithmic errors.

\begin{table}[htbp]
\centering
\setlength{\tabcolsep}{4pt}
\renewcommand{\arraystretch}{1.1}
\begin{tabular}{c c c c}
\hline
$\beta$ & $\sigma$ & $\beta_M (\delta = 1)$ & $\beta_M (\delta = 2)$ \\
\hline\hline
$1.00$ & $0.5$ & $3.9888(59)$ & $3.9941(64)$ \\
$1.00$ & $1.0$ & $1.9953(29)$ & $1.9980(36)$ \\
$1.00$ & $2.0$ & $0.9983(14)$ & $0.9994(22)$ \\
$1.00$ & $4.0$ & $0.4996(7)$  & $0.4986(14)$ \\
\hline
\end{tabular}
\caption{Estimated configurational temperature $\beta_M$ as a function of noise variance $\sigma$ at $\beta = 1.00$.}
\label{tab:lat_ptsymm_beta_algo_err}
\end{table}

\subsection{Dependence on Langevin step size}
\label{sec:Dependence_on_Langevin_step_size}

We next study the effect of the Langevin step size $\epsilon$.  
Although correct results are recovered as $\epsilon \to 0$, finite step sizes introduce discretization errors.  
Table~\ref{tab:fixed_step_size} shows that $\beta_M$ deviates increasingly from the target value as $\epsilon$ grows, with convergence breaking down at large $\epsilon$ (e.g., for $\delta = 2$ at $\epsilon \gtrsim 0.1$).  
The estimator thus provides a quantitative criterion for selecting a step size small enough to control discretization artifacts, complementing standard observable-based checks.

\begin{table}[htbp]
\centering
\setlength{\tabcolsep}{4pt}
\renewcommand{\arraystretch}{1.1}
\begin{tabular}{c c c c}
\hline
$\beta$ & $\epsilon$ & $\beta_M (\delta = 1)$ & $\beta_M (\delta = 2)$ \\
\hline\hline
$1.00$ & $0.0001$ & $1.0002(15)$ & $0.9845(22)$ \\
$1.00$ & $0.001$  & $0.9983(14)$ & $0.9994(22)$ \\
$1.00$ & $0.01$   & $0.9830(15)$ & $0.9898(23)$ \\
$1.00$ & $0.03$   & $0.9474(14)$ & $0.9663(29)$ \\
$1.00$ & $0.1$    & $0.8251(12)$ & -- \\
$1.00$ & $0.2$    & $0.6437(11)$ & -- \\
\hline
\end{tabular}
\caption{Estimated configurational temperature $\beta_M$ as a function of the fixed Langevin step size $\epsilon$ at $\beta = 1.00$.}
\label{tab:fixed_step_size}
\end{table}

\subsection{Monitoring thermalization}
\label{sec:Monitoring_thermalization}

We also examine the behavior of the estimator during thermalization.  
Starting from the initial configuration $\phi_{\rm initial} = - i$, both physical observables and $\beta_M$ approach their equilibrium values as the system equilibrates.  

\begin{figure*}[htbp]
\centering
\includegraphics[width=2.5in]{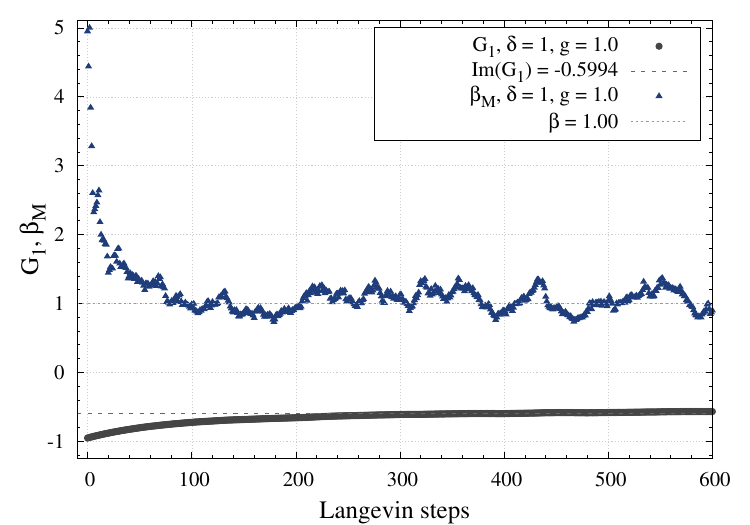}	
\includegraphics[width=2.5in]{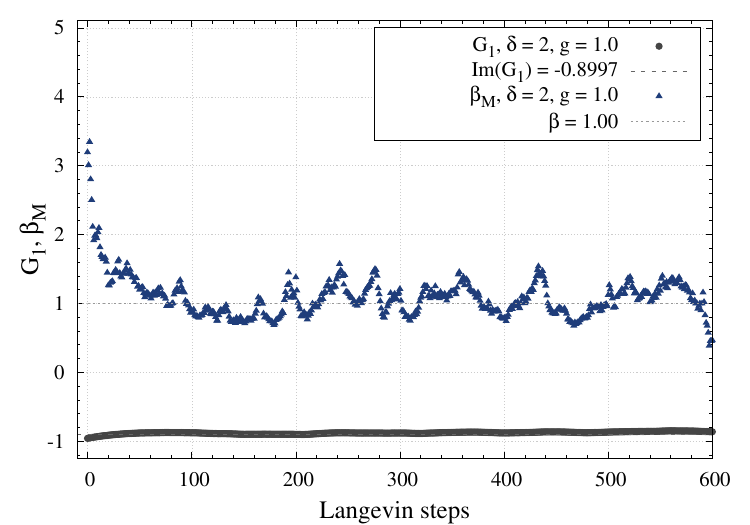}
\caption{Thermalization behavior of the one-point function $G_1 \equiv \langle \phi \rangle$ (imaginary part) and the configurational temperature estimator $\beta_M$ during the initial Langevin evolution. The data are for the PT-symmetric model with $\delta = 1$ (left) and $\delta = 2$ (right) at coupling $g = 1.0$.}
\label{fig:therm_behaviour}
\end{figure*}

Figure~\ref{fig:therm_behaviour} shows the Langevin-step evolution of the one-point function $G_1 = \langle \phi \rangle$ together with $\beta_M$ for $\delta = 1, 2$.  
Both quantities settle into stable distributions within $\sim200$ -- $300$ Langevin steps, with $\beta_M$ converging to the correct value $\beta = 1$ on the same timescale as $G_1$, demonstrating that the estimator provides a reliable monitor of thermalization.

\section{Comparison with Existing Diagnostics}
\label{sec:Comparison_with_Existing_Diagnostics}

\subsection{Langevin operator on observables}
\label{sec:Langevin_operator_on_observables}

\begin{figure*}[htbp]
\centering
\includegraphics[width=2.5in]{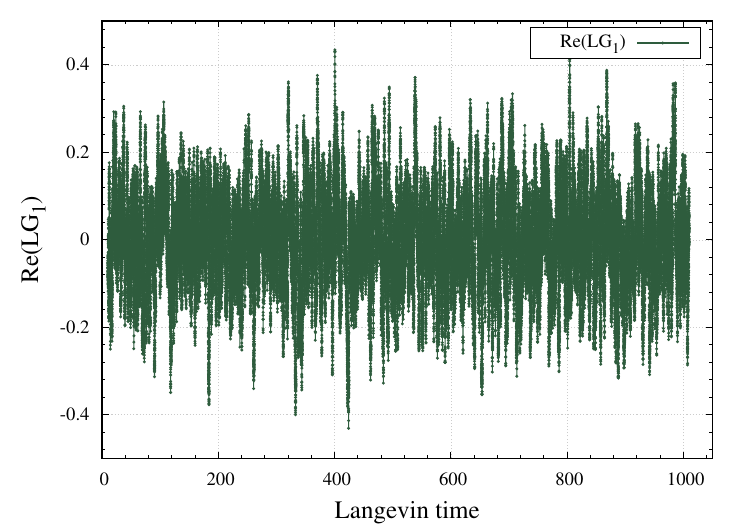}	
\includegraphics[width=2.5in]{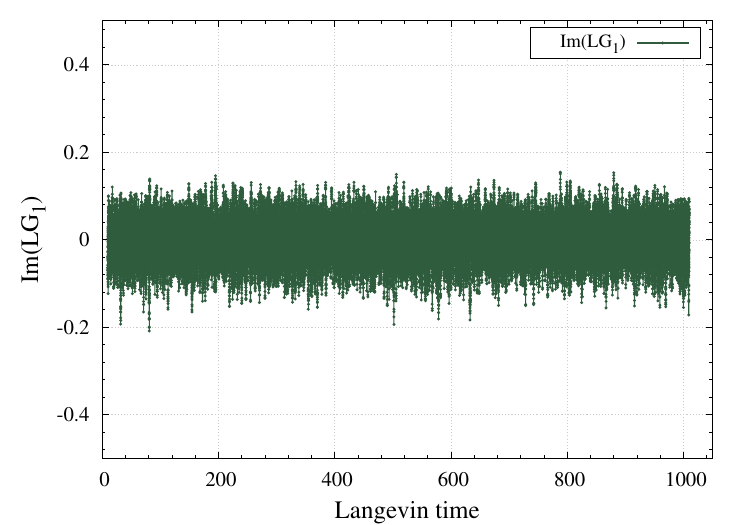}
\caption {The Langevin time history of $L G_1$. The data are for the PT-symmetric model with $\delta = 1$ and coupling $g = 1.0$.}
\label{fig:LG1_Lang_history}
\end{figure*}

A known correctness criterion for CLM is based on the Langevin operator acting on observables~\cite{Aarts:2011ax}.  
For an observable ${\cal O}_i[\phi, \tau]$,
\begin{equation}
\frac{\partial {\cal O}_i}{\partial\tau} = L_i \, {\cal O}_i, 
\qquad
L_i \equiv \left( \frac{\partial}{\partial\phi_i} - \frac{\partial \cal S}{\partial\phi_i} \right)\frac{\partial}{\partial\phi_i}.
\end{equation}
In equilibrium, one expects $\langle L_i {\cal O}_i \rangle = 0$. Applying this to the one-point function $G_1 = \langle \phi \rangle$ and averaging over sites, we find that $\langle L G_1\rangle$ is consistent with zero within errors (Fig.~\ref{fig:LG1_Lang_history}, Table~\ref{tab:comparison_three_criteria}), indicating that this criterion does not flag problems in the present simulations.

\subsection{Decay of the drift terms}
\label{sec:Decay_of_the_drift_terms}

A second criterion~\cite{Nagata:2016vkn} requires the probability distribution of the drift magnitude
\begin{equation}
u \equiv \sqrt{\frac{1}{N_\tau}\sum_{n = 0}^{N_\tau-1} \left|\frac{\partial S}{\partial\phi_n}\right|^2}
\end{equation}
to decay exponentially or faster at large $u$.

\begin{figure*}[htbp]
\centering
\includegraphics[width=2.5in]{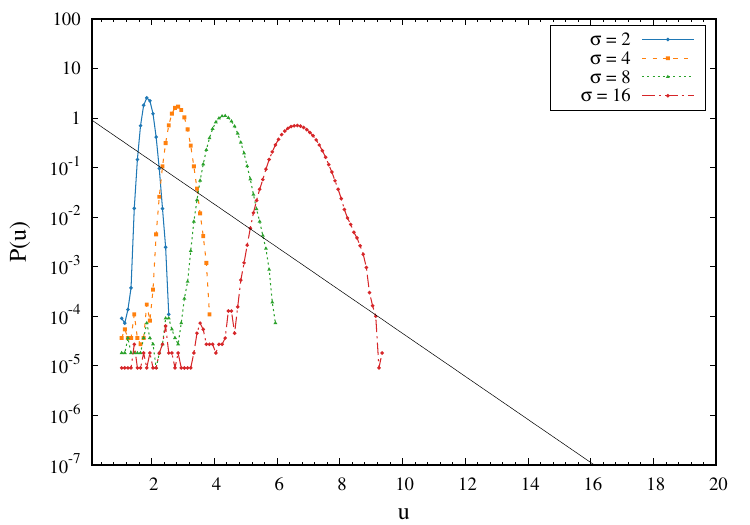}	
\includegraphics[width=2.5in]{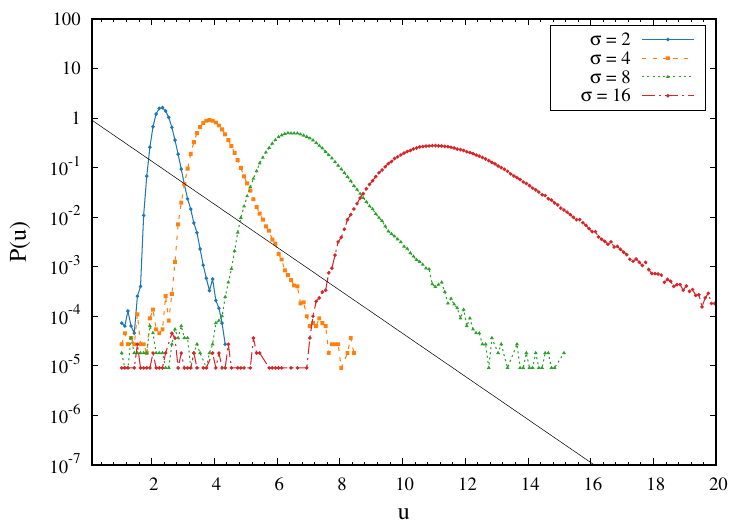}
\caption {Decay of the drift terms. The simulations are for the PT-symmetric model with $\delta = 1$ (left) and $\delta = 2$ (right) and coupling $g = 1.0$.}
\label{fig:Pu_u_g1_d1_d2_s_2_4_8_16}
\end{figure*}

Figure~\ref{fig:Pu_u_g1_d1_d2_s_2_4_8_16} shows the drift distributions $P(u)$ for several noise variances $\sigma$.  
For $\delta = 1$, exponential suppression holds for all tested $\sigma$.  
For $\delta = 2$, exponential decay is observed for $\sigma = 2, 4, 8$ but breaks down at $\sigma = 16$, signaling unreliable dynamics in that case.

\subsection{Comparing the three correctness criteria}
\label{sec:Comparing_the_three_correctness_criteria}

We summarize the three diagnostics as: $(i.)$ {\it Configurational temperature:} $|\beta_M - \beta| < 3 \sigma$ deviation, $(ii.)$ {\it Langevin operator:} $|\langle LG_1\rangle| < 3 \sigma$ deviation, and $(iii.)$ {\it Drift criterion:} exponential (or faster) decay of $P(u)$.

For correct noise normalization ($\sigma = 2$), all three criteria are satisfied.  
When controlled algorithmic errors are introduced via noise mis-scaling, only Criterion~(i) consistently detects the deviation.  
Criterion~(ii) remains satisfied in all cases, while Criterion~(iii) fails only for extreme mis-scalings.

The results are summarized in Table~\ref{tab:comparison_three_criteria}.  
We conclude that the configurational temperature provides a significantly sharper and more sensitive diagnostic, capable of detecting subtle algorithmic errors that remain invisible to existing CLM reliability criteria.

\begin{table}[htbp]
\centering
\setlength{\tabcolsep}{4pt}
\renewcommand{\arraystretch}{1.1}
\begin{tabular}{c c c c c c}
\hline
$\delta$ & $\beta$ & $\sigma$ &
Criterion~1 &
Criterion~2 &
Criterion~3 \\
\hline\hline
$1.0$ & $1.0$ & $2.0$  & $0.9983(14)$ (S) & $-0.0010(3)$ (S) & Exp.\ (S) \\
$1.0$ & $1.0$ & $4.0$  & $0.4984(2)$  (NS) & $-0.0014(5)$ (S) & Exp.\ (S) \\
$1.0$ & $1.0$ & $8.0$  & $0.2492(1)$  (NS) & $-0.0019(9)$ (S) & Exp.\ (S) \\
$1.0$ & $1.0$ & $16.0$ & $0.1246(0)$  (NS) & $-0.0027(14)$ (S) & Exp.\ (S) \\
\hline
$2.0$ & $1.0$ & $2.0$  & $0.9994(22)$ (S) & $-0.0010(5)$ (S) & Exp.\ (S) \\
$2.0$ & $1.0$ & $4.0$  & $0.4975(12)$ (NS) & $-0.0013(8)$ (S) & Exp.\ (S) \\
$2.0$ & $1.0$ & $8.0$  & $0.2485(3)$  (NS) & $-0.0018(15)$ (S) & Exp.\ (S) \\
$2.0$ & $1.0$ & $16.0$ & $0.1238(2)$  (NS) & $-0.0023(25)$ (S) & Non-exp.\ (NS) \\
\hline
\end{tabular}
\caption{Comparison of three correctness criteria for the PT-symmetric model at $g = 1.0$.
(S) and (NS) denote ``satisfied'' and ``not satisfied,'' respectively.}
\label{tab:comparison_three_criteria}
\end{table}

\section{Discussion and Outlook}
\label{sec:Discussion_and_Outlook}

We have proposed a configurational temperature estimator as a reliability diagnostic for the complex Langevin method. 
We used the one-dimensional PT-symmetric models as a test bed.
In these models, it reproduces the input temperature at the percent level and detects controlled algorithmic errors.
It also outperforms standard Langevin-operator and drift-based criteria in identifying noise mis-scaling (Tables~\ref{tab:lat-ptsymm-beta-n3-n4}, \ref{tab:lat_ptsymm_beta_algo_err}, \ref{tab:comparison_three_criteria}).

We note that the main advantage of the estimator is its direct test of thermodynamic consistency: $\beta_M$ probes whether configurations are sampled with the correct weight $e^{-{\cal S}[\phi]}$, rather than properties of the Langevin dynamics itself. 
This makes it a natural complement to existing CLM diagnostics and a universal correctness check, even when exact expectation values are unknown. 
We note that the estimator is local and not intrinsically dimension-dependent. 
Extending its application to higher-dimensional theories, and ultimately to lattice QCD at finite density, is a promising direction for future work.

\acknowledgments

We thank Navdeep Singh Dhindsa, Piyush Kumar, Vamika Longia, and Michael Mandl for their invaluable discussions. 
The work of A.J. was supported in part by a Start-up Research Grant from the University of the Witwatersrand. 
A.J. gratefully acknowledges the warm hospitality of the National Institute for Theoretical and Computational Sciences (NITheCS) and Stellenbosch University during the NITheCS Focus Area Workshop, {\it Decoding the Universe: Quantum Gravity and Quantum Fields.} 
The work of A.K. was partly supported by the National Natural Science Foundation of China under Grants No. 12293064, No. 12293060, and No. 12325508, as well as the National Key Research and Development Program of China under Contract No. 2022YFA1604900.

\bibliographystyle{JHEP}
\bibliography{lattice25}

@article{Joseph:2019sof,
    author = "Joseph, Anosh and Kumar, Arpith",
    title = "{Complex Langevin Simulations of Zero-dimensional Supersymmetric Quantum Field Theories}",
    eprint = "1908.04153",
    archivePrefix = "arXiv",
    primaryClass = "hep-th",
    doi = "10.1103/PhysRevD.100.074507",
    journal = "Phys. Rev. D",
    volume = "100",
    pages = "074507",
    year = "2019"
}

@article{Joseph:2020gdh,
    author = "Joseph, Anosh and Kumar, Arpith",
    title = "{Complex Langevin dynamics and supersymmetric quantum mechanics}",
    eprint = "2011.08107",
    archivePrefix = "arXiv",
    primaryClass = "hep-lat",
    doi = "10.1007/JHEP10(2021)186",
    journal = "JHEP",
    volume = "10",
    pages = "186",
    year = "2021"
}

@article{Kumar:2022fas,
    author = "Kumar, Arpith and Joseph, Anosh",
    title = "{Complex Langevin simulations for PT-symmetric models}",
    eprint = "2201.12001",
    archivePrefix = "arXiv",
    primaryClass = "hep-lat",
    doi = "10.22323/1.396.0124",
    journal = "PoS",
    volume = "LATTICE2021",
    pages = "124",
    year = "2022"
}

@article{Kumar:2022giw,
    author = "Kumar, Arpith and Joseph, Anosh and Kumar, Piyush",
    title = "{Complex Langevin Study of Spontaneous Symmetry Breaking in IKKT Matrix Model}",
    eprint = "2209.10494",
    archivePrefix = "arXiv",
    primaryClass = "hep-lat",
    doi = "10.22323/1.430.0213",
    journal = "PoS",
    volume = "LATTICE2022",
    pages = "213",
    year = "2023"
}

@article{Kumar:2023nya,
    author = "Kumar, Arpith and Joseph, Anosh and Kumar, Piyush",
    title = "{Investigating Spontaneous SO(10) Symmetry Breaking in~Type IIB Matrix Model}",
    eprint = "2308.03607",
    archivePrefix = "arXiv",
    primaryClass = "hep-lat",
    doi = "10.1007/978-981-97-0289-3_337",
    journal = "Springer Proc. Phys.",
    volume = "304",
    pages = "1201--1203",
    year = "2024"
}

@article{Joseph:2025tfw,
    author = "Joseph, Anosh and Kumar, Arpith",
    title = "{Complex Langevin simulations of supersymmetric theories}",
    eprint = "2504.02660",
    archivePrefix = "arXiv",
    primaryClass = "hep-lat",
    doi = "10.1142/S0217751X25300066",
    journal = "Int. J. Mod. Phys. A",
    volume = "40",
    number = "20",
    pages = "2530006",
    year = "2025"
}

@article{Klauder:1983nn,
author         = "Klauder, J. R.",
title          = "{Stochastic Quantization}",
booktitle      = "{Recent developments in high-energy physics. Proceedings,
22. Internationale Universitatswochen fur Kernphysik:
Schladming, Austria, February 23 - March 5, 1983}",
journal        = "Acta Phys. Austriaca Suppl.",
volume         = "25",
year           = "1983",
pages          = "251-281",
doi            = "10.1007/978-3-7091-7651-1_8",
reportNumber   = "PRINT-83-0321 (BTL)",
SLACcitation   = "%%CITATION = APAUA,25,251;%%"
}

@article{Parisi:1984cs,
author         = "Parisi, G.",
title          = "{On Complex Probabilities}",
journal        = "Phys. Lett.",
volume         = "131B",
year           = "1983",
pages          = "393-395",
doi            = "10.1016/0370-2693(83)90525-7",
SLACcitation   = "%%CITATION = PHLTA,131B,393;%%"
}

@article{Basu:2018dtm,
author         = "Basu, Pallab and Jaswin, Kasi and Joseph, Anosh",
title          = "{Complex Langevin Dynamics in Large $N$ Unitary Matrix
Models}",
journal        = "Phys. Rev.",
volume         = "D98",
year           = "2018",
number         = "3",
pages          = "034501",
doi            = "10.1103/PhysRevD.98.034501",
eprint         = "1802.10381",
archivePrefix  = "arXiv",
primaryClass   = "hep-th",
SLACcitation   = "%%CITATION = ARXIV:1802.10381;%%"
}

@article{Anagnostopoulos:2017gos,
author         = "Anagnostopoulos, Konstantinos N. and Azuma, Takehiro and
Ito, Yuta and Nishimura, Jun and Papadoudis, Stratos
Kovalkov",
title          = "{Complex Langevin analysis of the spontaneous symmetry
breaking in dimensionally reduced super Yang-Mills
models}",
journal        = "JHEP",
volume         = "02",
year           = "2018",
pages          = "151",
doi            = "10.1007/JHEP02(2018)151",
eprint         = "1712.07562",
archivePrefix  = "arXiv",
primaryClass   = "hep-lat",
reportNumber   = "KEK-TH-2023",
SLACcitation   = "%%CITATION = ARXIV:1712.07562;%%"
}

@article{Aarts:2009uq,
author         = "Aarts, Gert and Seiler, Erhard and Stamatescu,
Ion-Olimpiu",
title          = "{The Complex Langevin method: When can it be trusted?}",
journal        = "Phys. Rev.",
volume         = "D81",
year           = "2010",
pages          = "054508",
doi            = "10.1103/PhysRevD.81.054508",
eprint         = "0912.3360",
archivePrefix  = "arXiv",
primaryClass   = "hep-lat",
SLACcitation   = "%%CITATION = ARXIV:0912.3360;%%"
}

@article{Aarts:2011ax,
author         = "Aarts, Gert and James, Frank A. and Seiler, Erhard and
Stamatescu, Ion-Olimpiu",
title          = "{Complex Langevin: Etiology and Diagnostics of its Main
Problem}",
journal        = "Eur. Phys. J.",
volume         = "C71",
year           = "2011",
pages          = "1756",
doi            = "10.1140/epjc/s10052-011-1756-5",
eprint         = "1101.3270",
archivePrefix  = "arXiv",
primaryClass   = "hep-lat",
reportNumber   = "MPP-2011-3",
SLACcitation   = "%%CITATION = ARXIV:1101.3270;%%"
}

@article{Nagata:2016vkn,
author         = "Nagata, Keitaro and Nishimura, Jun and Shimasaki, Shinji",
title          = "{Argument for justification of the complex Langevin
method and the condition for correct convergence}",
journal        = "Phys. Rev.",
volume         = "D94",
year           = "2016",
number         = "11",
pages          = "114515",
doi            = "10.1103/PhysRevD.94.114515",
eprint         = "1606.07627",
archivePrefix  = "arXiv",
primaryClass   = "hep-lat",
reportNumber   = "KEK-TH-1911",
SLACcitation   = "%%CITATION = ARXIV:1606.07627;%%"
}

@article{PhysRevLett.78.772,
  title = {Dynamical Approach to Temperature},
  author = {Rugh, Hans Henrik},
  journal = {Phys. Rev. Lett.},
  volume = {78},
  issue = {5},
  pages = {772--774},
  numpages = {0},
  year = {1997},
  month = {Feb},
  eprint = "chao-dyn/9701026",
  archivePrefix = "arXiv",
  publisher = {American Physical Society},
  doi = {10.1103/PhysRevLett.78.772},
  url = {https://link.aps.org/doi/10.1103/PhysRevLett.78.772}
}

@article{10.1063/1.477301,
    author = {Butler, B. D. and Ayton, Gary and Jepps, Owen G. and Evans, Denis J.},
    title = {Configurational temperature: Verification of Monte Carlo simulations},
    journal = {The Journal of Chemical Physics},
    volume = {109},
    number = {16},
    pages = {6519-6522},
    year = {1998},
    month = {10},
    issn = {0021-9606},
    doi = {10.1063/1.477301},
    url = {https://doi.org/10.1063/1.477301}
}

@inproceedings{Mandl:2025ins,
    author = "Mandl, Michael and Seiler, Erhard and Sexty, D{\'e}nes",
    title = "{Complex Langevin simulations with a kernel}",
    booktitle = "{42th International Symposium on Lattice Field Theory}",
    eprint = "2512.14153",
    archivePrefix = "arXiv",
    primaryClass = "hep-lat",
    month = "12",
    year = "2025"
}
\end{document}